\documentclass[prl,aps,showpacs,twocolumn]{revtex4}

\usepackage{times}
\usepackage{graphicx}
\usepackage{epsfig}

\begin{document}


\title{A Duality in Entanglement Enabling a Test of Quantum Indistinguishability Unaffected by Interactions}

\author{S. Bose$^{1}$ and D. Home$^2$}
\affiliation{$^1$Department of Physics and Astronomy, University
College London, Gower St., London WC1E 6BT, UK}
\affiliation{$^2$CAPSS, Physics Department, Bose Institute, Salt Lake, Sector V, Kolkata
700097, India}







\begin{abstract}
We point out an earlier unnoticed implication of quantum
indistinguishability, namely, a property which we call `dualism' that characterizes the
entanglement of two
   identical particles (say, two ions of the same species) -- a feature which is absent in the entanglement of two
   non-identical particles (say, two ions of different species). A crucial application of this property is that it can be used to test
quantum indistinguishability without bringing the relevant particles together, thereby avoiding the effects of mutual interaction. This is in
contrast to the existing tests of quantum indistinguishability. Such a scheme, being independent of the nature and strength of mutual
interactions of the identical particles involved, has potential applications, including the probing of the transition from quantum
indistinguishability to classical distinguishability.
\end{abstract}

\maketitle




  A profound feature of the quantum world is
  the indistinguishability of various copies of a given particle -- a property which has been verified to hold for photons \cite{HBT,sandoghdar}, mesons \cite{baym}, electrons \cite{yamamoto98}, neutrons \cite{ianuzzi} and recently for He and Rb atoms \cite{aspect,bloch,Rbexpt}. The last set of experiments are significant attempts in testing quantum indistinguishability (QI) for increasingly massive objects. We may stress that extending the verification of QI to objects more complex than atoms is of fundamental importance since this will probe the limits of the quantum world \cite{leggett-PhyScr} from a perspective which is {\em distinct} from testing the superposition principle for macroscopic systems -- a
widely pursued program \cite{leggett-PhyScr,macro} that has notably advanced last year \cite{martinis-cleland-expt}. A further motivation comes from a recent study of tunably indistinguishable photons  \cite{sandoghdar} that leads to the question whether similar tunings of QI can occur while verifying it with increasingly macroscopic objects. A key condition for such tests \cite{yamamoto98,aspect,sandoghdar,bose-home} is to ensure that the observed statistical
effects arise {\em solely} from QI. This requirement is hard to satisfy for macroscopic or any other type of strongly interacting objects, such as while testing the QI of large molecules by bringing them together at a beam splitter. For example, for mutually repelling bosonic objects, fermionic behaviour may be seen \cite{stenholm}. In this context, the very recent photonic simulations of the effects of interactions on the tests of QI \cite{nat-phot} underscore the topical interest of this issue \cite{Rbexpt,bloch}. Thus, the question arises whether
QI can be tested in a way that is {\em unaffected} by mutual interactions of the objects involved. That such an `interaction-independent test' of QI is indeed possible is revealed by the present work. This possibility arises from a hitherto unexplored property of an entangled
state of identical particles which we call `dualism'.

The above mentioned `dualism' can be stated as follows: If two identical particles (IPs), distinctly labeled by a dynamical
variable ${\cal A}$, are entangled in terms of a different dynamical variable ${\cal B}$, then these particles can also be regarded as being
entangled in the variable ${\cal A}$ when labeled by the other variable ${\cal B}$. Interestingly, this feature provides a testable difference
between an entangled state of IPs, say, two ions of the same species localized in traps at distinct locations (as in Ref.\cite{monroe}), and an entangled state of non-identical
particles (NIPs), say, an ion and a photon (as in Ref.\cite{duan}). Here we formulate a generic scheme to {\em
test} this dualism that is implementable with any pair of distinctly labelled IPs, provided they can be entangled. While such entangled states are routinely produced for photons \cite{bouwmeester} and
trapped ions \cite{monroe}, very recently, the productions of entangled mobile electrons \cite{schoenberger} and trapped atoms \cite{rydberg}, along with notable advances in entangling mobile atoms \cite{twin} and distant molecules \cite{sandoghdar} have been achieved. Here an important point is that if QI of given IPs is to be tested, this property should
not in itself be invoked to generate the required entanglement. From this point of view, the entangling mechanisms of Refs.
\cite{sandoghdar,monroe,rydberg,micheli,rabl,milman} are particularly apt. Further, we may note that the scheme proposed here could be practically useful. For
example, it implies that a given entangled state of spin/internal degrees of freedom
\cite{bouwmeester,monroe,schoenberger,rydberg,micheli,rabl,milman} can also function as a momentum entangled state. It may thus allow the
flexibility of invoking the {\em same} entangled state as a resource for processing quantum information using either the
internal or the motional variables.

   While the present paper will be couched througout in terms of entangled "particles", the treatment will be equally valid in terms
of entangled "modes" \cite{mode} when each mode has {\em exactly one} particle \cite{vaccaro}. Situations abound in which two identical particles can be distinctly labelled using a suitable
dynamical variable- say, the EPR-Bohm type states of two identical particles where the terminology particle 1 and 2 is widely
used. Such distinct identification may be made through a difference in spatial locations of the particles (such as ions in distinct traps), or in their momenta (such as photons flying in different directions). These types of entangled states are crucial in applications of quantum information, where the terminology such as
"a local operation on particle 1" (say, belonging to a party Alice) and "a local operation on particle 2"
(say, belonging to a party Bob) is frequently used even if the particles under consideration (e.g. two photons or two electrons)
are identical. On the other hand, the two correlated electrons in a Helium atom exhibiting quantum indistinguishability cannot be distinctly labelled - however, it is important to stress that in our paper we are {\em not} considering such a situation. In the present paper, we proceed to show that in the former EPR-Bohm type situation, by formulating a suitable example, quantum indistinguishability can be
made to manifest for identical particles that are distinctly labelled. This becomes possible because of the way the choice of the dynamical variable for labeling the particles is appropriately varied in the course of our experimental scheme.

Let us
consider the EPR-Bohm entangled state of two spin-$\frac{1}{2}$
  IPs (e.g. electrons) written as
\begin{equation}
|\Psi\rangle_{12}=\alpha|\uparrow\rangle_1|\downarrow\rangle_2+\beta|\downarrow\rangle_1|\uparrow\rangle_2
\label{bohm}
\end{equation}
where $\alpha$ and $\beta$ are non-zero amplitudes. In writing
Eq.(1), the labels $1$ and $2$ need to correspond to different
values of dynamical variables because the identical particles cannot
be distinguished in terms of their innate attributes such as rest
mass or charge. Although Eq.(\ref{bohm}) is widely used, an alternative description of the above EPR-Bohm
      entangled state of IPs is given in the usual second quantized notation as $|\Psi\rangle_{12}=(\alpha~
c^\dagger_{k_1,\uparrow}c^\dagger_{k_2,\downarrow}+\beta~
c^\dagger_{k_1,\downarrow}c^\dagger_{k_2,\uparrow})|0\rangle$, where $c^\dagger_{k_i,\sigma}$ creates a particle in momenta state $k_i$ and spin state $\sigma=\uparrow,\downarrow$ and $|0\rangle$ is the vacuum state. This second quantized representation clarifies that in order to meaningfully describe an EPR-Bohm state of
two identical particles we need {\em at least two variables}: One variable
${\cal A}$ (e.g. momentum in the above example) to label the particles, and
another variable ${\cal B}$ (e.g. spin) which is entangled, where $[{\cal
A},{\cal B}]=0$. In terms of distinct eigenvalues $A_1,A_2$ and $B_1,B_2$ of the variables ${\cal A}$ and ${\cal B}$ respectively, one may
thus write an EPR-Bohm state as
\begin{equation}
|\Psi(A_1,A_2,B_1,B_2)\rangle=(\alpha~
c^\dagger_{A_1,B_1}c^\dagger_{A_2,B_2}+\beta~
c^\dagger_{A_1,B_2}c^\dagger_{A_2,B_1})|0\rangle,
\label{second}
\end{equation}
where $c^\dagger_{A_i,B_j}$ creates a particle in the simultaneous
eigenstate $|A_i,B_j\rangle$ of the variables ${\cal A}$ and ${\cal
B}$.  In order to put Eq.(\ref{second}) in the form of Eq.(\ref{bohm}) we {\em rewrite}
$c^\dagger_{A_i,B_j}|0\rangle$ as $|B_j\rangle_{A_i}$ where  $A_i$ is taken as the ``which-particle" label (thus, $|B_j\rangle_{A_i}$ is
a second quantized notation), whence
\begin{equation}
|\Psi(A_1,A_2,B_1,B_2)\rangle=\alpha~|B_1\rangle_{A_1}|B_2\rangle_{A_2}+
\beta~|B_2\rangle_{A_1}|B_1\rangle_{A_2}
 \label{state0}
\end{equation}
The above form of {\em rewriting} is, in fact, standard and is widely used in describing the
entangled states of IPs generated in actual/proposed experiments
\cite{bouwmeester,bose-home,gerry,egues}. For example, in the
routinely used two-photon entangled state
$\frac{1}{\sqrt{2}}(|H\rangle_1|V\rangle_2+|V\rangle_1|H\rangle_2)$,
the symbols $|H\rangle_i$ and $|V\rangle_j$ are, in fact, rewritten
forms for $c^{\dagger}_{k_i,H}|0\rangle$ and
$c^{\dagger}_{k_j,V}|0\rangle$, where $k_i$ and $k_j$ are labels for
distinct momenta directions.
 Considering in the same sense as
Eq.(\ref{bohm}), the above Eq. (\ref{state0}) is an entangled state
of the variable ${\cal B}$ (say, polarization or spin) with the
variable ${\cal A}$ (say, position or momentum) being the ``which
particle" label. {\em Alternatively}, we may use the eigenvalues of
the variable ${\cal B}$ to label the particles and replace
$c^\dagger_{A_i,B_j}|0\rangle$ in Eq.(\ref{second}) by
$|A_i\rangle_{B_j}$ to {\em rewrite} Eq.(\ref{second}) as
\begin{equation}
|\Psi(A_1,A_2,B_1,B_2)\rangle=\alpha~|A_1\rangle_{B_1}|A_2\rangle_{B_2}\pm \beta~ |A_2\rangle_{B_1}|A_1\rangle_{B_2}\label{state1}
\end{equation}
where, in the last step, $|A_2\rangle_{B_1}\equiv c^\dagger_{A_2,B_1}$ and $|A_1\rangle_{B_2}\equiv c^\dagger_{A_1,B_2}$ have been exchanged to bring Eq.(\ref{state1}) to the same form as Eq.(\ref{bohm}) (i.e., the ``which particle" label $B_1$
 preceding the ``which particle" label $B_2$ in both terms of the superposition). The upper and lower signs of $\pm$ in
Eq.(\ref{state1}) correspond to bosons and fermions respectively and arises from the above exchange of creation operators. The two equivalent representations of the state $|\Psi(A_1,A_2,B_1,B_2)\rangle$ given by Eqs.(\ref{state0})
    and (\ref{state1}) bring out the property of dualism. This means that a
class of states of two identical particles can {\em equally} well be regarded as entangled in {\em either} the variable ${\cal A}$ {\em or} the
variable ${\cal B}$, depending upon whether the variable used for distinguishing (labelling) the particles is ${\cal B}$ {\em or} ${\cal A}$
respectively. The way this property of dualism arises can also be seen clearly through the derivation given in the {\em supplementary material} \cite{suppl} in terms of  a first quantized formulation based on appropriate symmetrization/anti-symmetrization using pseudo-labels. 

 That the above property of dualism holds essentially for IPs can be seen by replacing for NIPs,  the right
hand side of Eq.(\ref{second}) by $(\alpha~ c^\dagger_{A_1,B_1}d^\dagger_{A_2,B_2}+\beta~ c^\dagger_{A_1,B_2}d^\dagger_{A_2,B_1})|0\rangle$
where $c^\dagger$ and $d^\dagger$ create different species of particles. While the above state can be written in the analogue of Eq.(\ref{state0}):  $\alpha~|B_1\rangle_{A_1,C}|B_2\rangle_{A_2,D}+
\beta~|B_2\rangle_{A_1,C}|B_1\rangle_{A_2,D}$, where $C$ and $D$ stand for distinct particle attributes such as mass/charge, it cannot be written in the analogue of Eq.(\ref{state1}), as that would entail superposing states $|C\rangle$ and $|D\rangle$ which is not allowed.

There is a {\em complementarity} in the dualism in the sense that one cannot
 observe simultaneously the entanglement in {\em both} the variables. This complementarity makes evident the
way $|\Psi(A_1,A_2,B_1,B_2)\rangle$ differs from the hyper-entangled states \cite{kwiat} in which more than one variable is simultaneously
entangled. We may also stress that this property of dualism, stemming from the {\em interchangeability} of two different dynamical variables that are used for labeling the concerned particles, is a manifestation of quantum indistinguishability that is different from its other manifestations, e.g. the behaviour of IPs on simultaneous incidence on a 
beam-splitter
   \cite{bouwmeester}.

\begin{figure}
\begin{center}
\leavevmode \epsfxsize=7cm \epsfbox{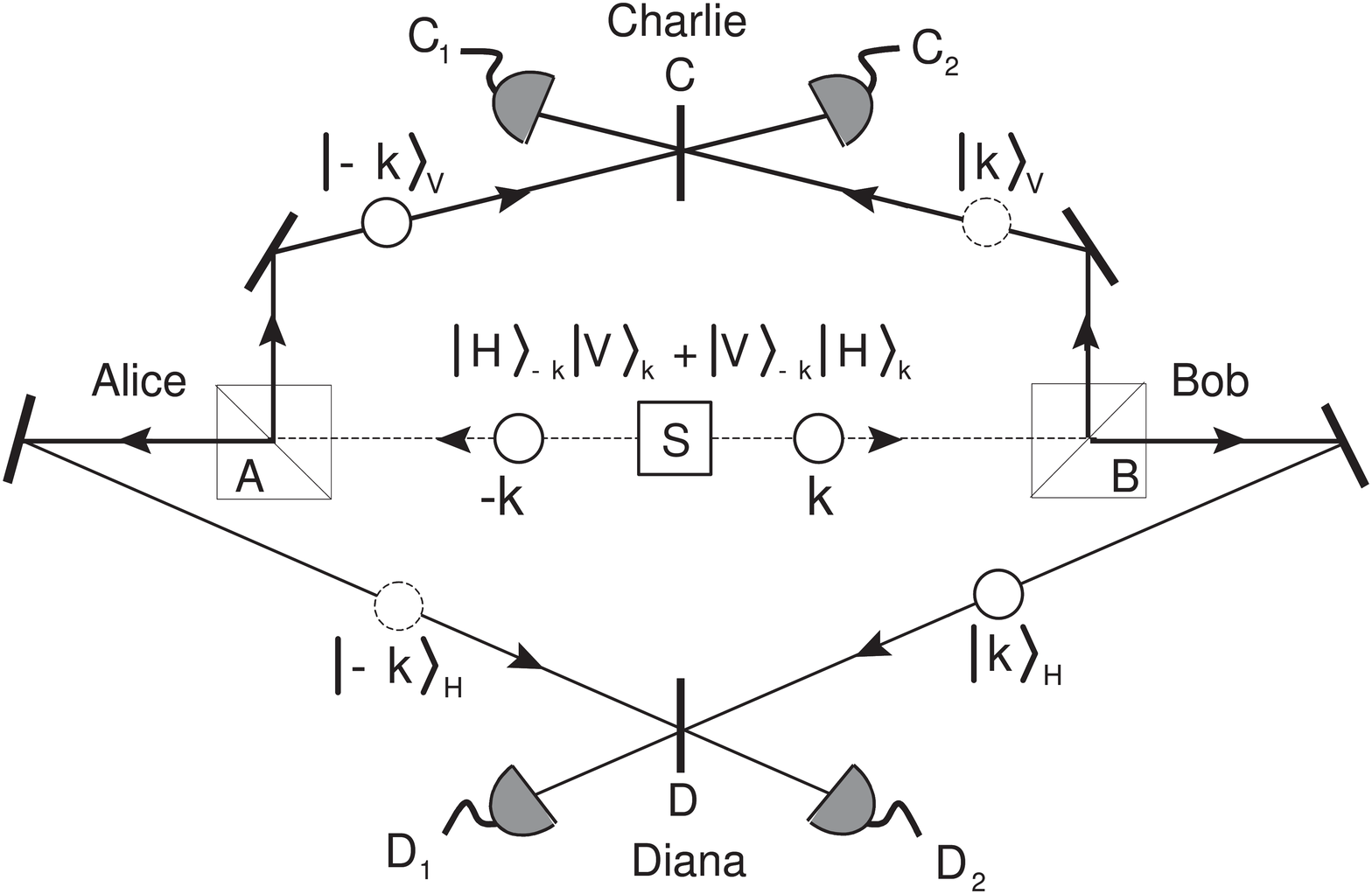}
 \caption{Scheme to test the dualism in entanglement of identical particles.
 Two photons of the same frequency in the polarization entangled state $|\Psi\rangle_{12}=\frac{1}{\sqrt{2}}(|H\rangle_{1}|V\rangle_{2}+|V\rangle_{1}|H\rangle_{2})$ are emitted in opposite directions by the
source S. The labels $1$ and $2$  stand for their momenta ${\bf -k}$
and  ${\bf k}$ respectively along the $x$ axis. The ${\bf -k}$ and
${\bf k}$ photons fly towards
 Alice and Bob respectively. If they conducted
 polarization correlation measurements on these photons they could verify their polarization
 entanglement. Instead, in order to verify the {\em dual} momentum entanglement implied by the right hand side
 of Eq.(\ref{dualexp}), they put polarization
 beam splitters A and B in the paths of the photons which deflects
 $|V\rangle$ photons
 in the $+y$ direction  (towards Charlie), and $|H\rangle$ photons in the $-y$ direction (towards Diana). It is because the
 photon pair is emitted in the state $|\Psi\rangle_{12}$ that only one photon reaches Charlie, while its partner reaches Diana. Then, by the dualism of Eq.(\ref{dualexp}), the $x$-component of their momenta is
entangled. Now, since there are only two possible values of
 the
 $x$-component of the momentum, namely ${\bf -k}$ and ${\bf k}$, one can associate a dichotomic pseudo-spin
 observable \cite{home} with this momentum using operators $\sigma_x=|{\bf -k}\rangle\langle {\bf k}|+|{\bf k}\rangle\langle {\bf -k}|,\sigma_y=i(|{\bf -k}\rangle\langle {\bf k}|-|{\bf k}\rangle\langle {\bf -k}|)$ and
 $\sigma_z=|{\bf k}\rangle\langle {\bf k}|-|{\bf -k}\rangle\langle {\bf -k}|$. These pseudo-spin operators and their linear combinations
 are
 measurable
 by a beam splitter with a tunable reflectivity and two detectors if exactly one photon is incident on the beam splitter
 at a time. By noting the
 coincidence of clicks in their detectors, Charlie and
 Diana can verify whether a Bell-type inequality is violated by the
 momentum pseudo-spin correlation measurements on their photons,
 thereby testing the property of dualism.}

\label{dual1}
\end{center}
\end{figure}

     Next, we discuss how the above dualism can be tested. Since an entangled
state is required for this purpose, we first consider the readily
available polarization entangled states of photons. Such a state can
be written as in Eq.(\ref{bohm}) with the particle indices $1$ and
$2$ corresponding to momenta labels ${-\bf k}$ and ${\bf k}$
respectively, and $\uparrow$ and $\downarrow$ representing
polarization states $H$ and $V$ respectively. The dualism can then
be expressed as
\begin{equation}
|H\rangle_{\bf -k}|V\rangle_{\bf k}+|V\rangle_{\bf
-k}|H\rangle_{\bf k}\equiv |{\bf -k}\rangle_{H}|{\bf
k}\rangle_{V}+ |{\bf k}\rangle_{H}|{\bf -k}\rangle_{V}.
\label{dualexp}
\end{equation}
Let ${\bf k}$ be chosen to be along the $x-$axis. Then the
polarization entanglement implied by the left hand side of
Eq.(\ref{dualexp}) can be tested in the usual manner by Alice and
Bob on opposite locations along the $x-$axis. For testing its dual,
we separate the $H$ and $V$ components of the state along the
$y-$axis as shown in Fig.\ref{dual1} with the aid of a polarization
beam splitter (PBS). Then the labels $H$ and $V$ on the right hand
side of Eq.(\ref{dualexp}) become identifiable with distinct momenta
along the $y$ axis, and the particle labeled as $H$ reaches Charlie,
while the particle labeled as $V$ reaches Diana as shown in
Fig.\ref{dual1}. Charlie and Diana are thus in possession of the
entangled state $|{\bf -k}\rangle_{C}|{\bf k}\rangle_{D}+ |{\bf
k}\rangle_{C}|{\bf -k}\rangle_{D}$, where $C$ stands for the
particle possessed by Charlie and $D$ for the particle possessed by
Diana. In this entangled state, the momenta component along the $x-$
axis appears as a simple dichotomic variable on which Charlie and
Diana can perform a Bell's inequality experiment using a
beam-splitter (BS) and detectors using the procedure described in
detail in the caption of Fig.\ref{dual1}. We stress that the violation of Bell's inequality by the dual forms of entanglement is merely a convenient tool to verify the property of
dualism (an {\em entanglement witness} can also be used \cite{witness}). Importantly, the same test is also possible with other entangled IPs such as ions
\cite{monroe,rowe} where efficient detectors make the study of Bell's inequality free of the detection loophole \cite{rowe}. Moreover, in this
context, the necessity of ensuring space-like separation does not arise at all.

The predicted {\em sign difference} between the dual forms of entanglement Eqs.(\ref{state0}) and (\ref{state1}) in the case of fermions should
also be testable. The presence of such sign difference is {\em reinforced} through an alternative derivation of the dualism in terms of symmetrization/antisymmetrization in the first quantized notation, given in the supplementary material \cite{suppl}. If separate experiments measuring the expectation value of the Bell operator (i.e. the expectation value of the
linear combination of four correlators occurring on the left hand side of the Bell inequality, without taking the modulus)  are performed using entanglements in the variables ${\cal
A}$ and ${\cal B}$ respectively, the expectation values in the two cases will have an opposite sign for fermions. Hence, the testing of such dualism can enable verifying the bosonic/fermionic nature of the particles.

 Further, note
that the PBS in the above scheme is merely used to separate the photons according to their polarization and enable the verification of the dual
entanglement (the PBS has {\em no role} in creating the dual entanglement). A practical application of the above scheme would thus be to use
spin/polarization entangled states (if they are easier to produce) as a {\em resource} for obtaining momentum entangled states. For example,
spin entangled mobile electrons have just been realized \cite{schoenberger}, where the state $|\uparrow\rangle_{\bf
k_1}|\downarrow\rangle_{\bf k_2}+|\downarrow\rangle_{\bf k_1}|\uparrow\rangle_{\bf k_2}$ is produced with ${\bf k_1}$ and ${\bf k_2}$ being
momenta states in two distinct one dimensional conducting channels. When such a state is generated, the dualism pointed out will enable one to
easily obtain a momentum entangled state (electronic wave-guides have been fabricated \cite{kasevich} and spin analogues of a PBS have been proposed \cite{foldi}). Almost any other method for
obtaining momenta entangled states {\em from} spin entangled states will be more complicated (involving either delocalized spin measurements or
additional momenta dependent spin flips). Next, we discuss a foundational application of the property of dualism.

Existing tests of QI involve
   bringing
   two IPs together \cite{HBT,yamamoto98,bose-home} to exhibit, for example, {\em bunching} and {\em anti-bunching}, whose results would be modified by particle interactions \cite{stenholm}.
   As the identical objects get complex, the
outcomes of such tests could increasingly deviate from the ideal non-interacting case in view of complicated scatterings, including inelastic collisions/fragmentations/chemical reactions. However,
in testing our dualism as above, the objects in the $|{\bf k}\rangle$ and the $|{\bf -k}\rangle$ state are {\em never} present concomitantly.
Consequently, the outcomes of our proposed experiment should be the {\em same} whatever be the mutual interactions of the IPs, thereby providing
an {\em interaction independent} test of QI.

    The caveat is that the IPs will have to be in an entangled state {\em before} the experiment. Preparing such a state is generically
challenging, for which invoking QI could be required \cite{bose-home}. However, there are {\em several} schemes which do not invoke QIs of
the IPs involved -- e.g., in a recent breakthrough atoms held in distinct tweezers were entangled \cite{rydberg}. The internal levels of two identical ions in distant traps have already been entangled by photo-detection
\cite{monroe}, whose feasibility for molecules has also been demonstrated last year \cite{sandoghdar}. In fact, the generation of entangled molecules is being
explored so actively that testing Bell's inequalities with molecules is just a matter of time \cite{milman}. Polar molecules (e.g.,
$CO,ND_3,OH,YbF$) cooled to a ground state and trapped \cite{meijer},  can be entangled either by a direct interaction \cite{micheli}, or
through a mediating resonator \cite{rabl}. Although the interactions are used here to generate an entangled state which is used to test QI
through our dualism, the test of QI in itself remains unaffected by interactions, provided the entangled particles are kept well separated
during the test.

    The entangling methods of the previous paragraph generate the
    state
$|\uparrow\rangle_{\psi_1}|\downarrow\rangle_{\psi_2}+|\downarrow\rangle_{\psi_1}|\uparrow\rangle_{\psi_2}$ where $\uparrow/\downarrow$ stand
for ionic/molecular internal states/spins, $\psi_1$ and $\psi_2$ label the center of mass (COM) wavefunction of ions/molecules in the traps $1$
and $2$ respectively, and $|\uparrow\rangle_{\psi_j}\equiv c^{\dagger}_{\uparrow,\psi_j}$ and $|\downarrow\rangle_{\psi_j}\equiv
c^{\dagger}_{\downarrow,\psi_j}$ (the rotational-vibrational modes of the molecules are taken to be cooled to their ground states
\cite{meijer,micheli}). To verify the dualism using our scheme (Fig.\ref{dual1}), one needs to transfer the entangled ions/molecules
from their traps to matter waveguides \cite{schmeidmyer}, thereby converting their trap states to momenta states: $|\psi_1\rangle\rightarrow
|{\bf -k}\rangle,~|\psi_2\rangle\rightarrow |{\bf k}\rangle$. For the subsequent interferometric procedure, beam-splitters/waveguides/PBS are
available (using atom-chips \cite{stenholm,schmeidmyer}, molecular wave-guides \cite{chinese,pbsmol} and molecule-chips \cite{santambrogio}). This opens up a way to reveal the true
indistinguishability of those mutually repulsive bosonic ions and polar molecules which show deceptive anti-bunching \cite{stenholm}.

We now consider the role in the dualism experiment of those degrees
of freedom which are not involved directly in any of the dual forms
of entanglement. These degrees of freedom are assumed to be in the
{\em same} collective state $|\chi_0\rangle$ for both the trapped
IPs when they are first entangled. We thus write the initial state
as
$|\uparrow\rangle_{\psi_1,\chi_0}|\downarrow\rangle_{\psi_2,\chi_0}+|\downarrow\rangle_{\psi_1,\chi_0}|\uparrow\rangle_{\psi_2,\chi_0}$.
 Now, suppose {\em after} the preparation of the above state and the switching off of any interaction between the IPs (say, by
 pulling their traps far apart), they are held in their respective traps for a
  time $t$ over which neither the COM motion nor the entangled spin-like variable is significantly affected
 by the environment. This is possible, since the spin-like variables considered by us can have long coherence times \cite{micheli} and
stable superpositions of motional states of the COM have been demonstrated for large molecules \cite{macro}. However, with increasing molecular
complexity, the number of degrees of freedom involved in $|\chi_0\rangle$ becomes larger with their collective state more influenced by the
environment. Then, $|\chi_0\rangle$ would evolve {\em differently} to $|\chi_1(t)\rangle$ and $|\chi_2(t)\rangle$ in the respective traps.
Although this does not affect the violation of Bell's inequality for the spin entanglement, it {\em suppresses} Bell's inequality violation for
the dual momentum entanglement by a factor $|\langle\chi_1(t)|\chi_2(t)\rangle|^2$ -- a form of decoherence relevant to the transition from QI
to classical distinguishability of IPs. In the classical limit, $\chi_1(t\rightarrow\infty)$ and $\chi_2(t\rightarrow\infty)$ emerge as
intrinsic labels for IPs as $|\langle\chi_1(t\rightarrow\infty)|\chi_2(t\rightarrow\infty)\rangle|^2\rightarrow 0$. Such a quantum to classical transition complements the widely studied quantum to classical transition through the
decoherence of superpositions \cite{macro,caldiera}.

Finally, the very feature, as we have shown, that there is a scheme {\em
unaffected} by interactions whereby one can test whether strongly interacting identical complex particles can `justly be regarded as being
created from the same vacuum' should be interestig in itself. Further, that this stems from a hitherto unnoticed dualism in the entanglement of IPs enhances the need for its experimental verification even for photons/ions/atoms/electrons. Also, importantly, as we have argued, if tested with more
complex objects, this dualism has the potentiality to provide a fruitful way of studying the transition from QI to classical distinguishability.

  We are grateful to A. J. Leggett and V. Singh for their useful remarks. DH acknowledges support from the Royal Society-Indian National Science Academy Exchange Programme which
initiated this collaboration, Centre for Science, Kolkata and the DST project. SB acknowledges support of the EPSRC UK, the Royal Society and
the Wolfson foundation.


\end{document}